\documentclass[aps,prl,twocolumn,amsmath,amssymb,showpacs]{revtex4-1}

\usepackage[utf8]{inputenc}
\usepackage{graphicx}
\usepackage{textcomp}
\usepackage{amsmath}
\usepackage{amssymb}
\usepackage{siunitx}
\usepackage{hyperref}
\usepackage{xcolor}

\begin{document}

\title{Spectral multiplexing of telecom emitters with stable transition frequency}
\author{Alexander Ulanowski}
\author{Benjamin Merkel}
\author{Andreas Reiserer}
\email{andreas.reiserer@mpq.mpg.de}

\affiliation{Max-Planck-Institut f\"ur Quantenoptik, Quantum Networks Group, Hans-Kopfermann-Strasse 1, D-85748 Garching, Germany}
\affiliation{Technical University of Munich, TUM School of Natural Sciences and Munich Center for Quantum Science and Technology (MCQST), James-Franck-Str. 1, D-85748 Garching, Germany}

\begin{abstract} 
In a quantum network, coherent emitters can be entangled over large distances using photonic channels. In solid-state devices, the required efficient light-emitter interface can be implemented by confining the light in nanophotonic structures. However, fluctuating charges and magnetic moments at the nearby interface then lead to spectral instability of the emitters. Here we avoid this limitation when enhancing the photon emission up to $70(12)$-fold using a Fabry-Perot resonator with an embedded $\SI{19}{\micro\meter}$ thin crystalline membrane, in which we observe around 100 individual erbium emitters. In long-term measurements, they exhibit an exceptional spectral stability of $<\SI{0.2}{\mega\hertz}$ that is limited by the coupling to surrounding nuclear spins. We further implement spectrally multiplexed coherent control and find an optical coherence time of $\SI{0.11+-0.01}{\milli\second}$, approaching the lifetime limit of $\SI{0.3}{\milli\second}$ for the strongest-coupled emitters. Our results constitute an important step towards frequency-multiplexed quantum-network nodes operating directly at a telecommunication wavelength.
\end{abstract} 

\maketitle

\section{Introduction}

Quantum networks can solve tasks and allow their users to interact in ways that are not possible using present-day technology \cite{wehner_quantum_2018}. Pioneering experiments have used atoms trapped in vacuum \cite{monroe_scaling_2013, reiserer_cavity-based_2015}, quantum dots \cite{lodahl_interfacing_2015}, and color centers in diamond \cite{ruf_quantum_2021}. To access the full potential of quantum networks, these prototypes need to be scaled to longer distances and larger qubit numbers. To this end, one needs to overcome the inefficiency and imperfections of photon transmission over large distances, which can be achieved by the techniques introduced in the seminal quantum repeater proposal \cite{briegel_quantum_1998}.

Implementing this and related protocols for distributed quantum computing \cite{nickerson_freely_2014, kinos_roadmap_2021} requires, first, an efficient emitter-photon interface, which can be achieved with optical resonators with small volume and large quality factor \cite{reiserer_cavity-based_2015,  lodahl_interfacing_2015, janitz_cavity_2020}. Second, to bridge large distances using optical fibers, the photons should be emitted in \cite{dibos_atomic_2018} or converted to \cite{zaske_visible--telecom_2012} the minimal-loss wavelength regime around $\SI{1550}{\nano\meter}$. Third, one needs to operate and control multiple qubits per quantum network node, e.g. by using nuclear spin registers \cite{bradley_ten-qubit_2019} or by spectral multiplexing \cite{chen_parallel_2020}. Finally, these qubits have to emit at a reproducible frequency, maintain their optical coherence during the photon emission, and their ground-state coherence until entanglement with a remote node is reliably established. Combining all of these properties in a single experimental platform is an outstanding challenge.

While efficient light-matter interfaces have been realized in several solid-state platforms using nanophotonic resonators \cite{lodahl_interfacing_2015, janitz_cavity_2020, dibos_atomic_2018, chen_parallel_2020, zhong_optically_2018, xia_tunable_2022}, in this approach the proximity of fluctuating charges and paramagnetic impurities at nearby interfaces leads to noise that is detrimental to the spectral stability of the emitters. Thus, except for a recent work in SiC \cite{lukin_optical_2022}, the cavity-enhanced generation of coherent photons has only been achieved with emitters that exhibit a low sensitivity to electric fields due to a zero first-order Stark coefficient that is a result of their symmetry \cite{macfarlane_optical_2007, thiering_ab_2018}. These emitters \cite{bhaskar_experimental_2020, zhong_optically_2018, kindem_control_2020} did not operate at a telecommunication wavelength and often required dilution refrigerators to ensure coherent operation. 

Here, we implement an alternative approach that requires neither insensitive emitters nor mK temperature as it avoids the proximity of interfaces. Thus, it should be applicable to a large variety of emitters and host materials. In our experiment, we place a thin crystalline membrane of erbium-doped Yttrium-Orthosilicate (Er:YSO) in a cryogenic Fabry-Perot resonator \cite{riedel_deterministic_2017} with a high quality factor of $10^7$. Compared to our earlier experiment on ensemble spectroscopy \cite{merkel_coherent_2020}, we use the same crystal and mirror, but a changed mechanical arrangement (see Materials and Methods). In spite of the considerable first-order Stark shift of Er:YSO, around $10\frac{\si{\kilo\hertz}}{\si{\volt\centi\meter}}$ \cite{jiri_minar_electric_2009}, we demonstrate coherent emission of single photons with exceptionally narrow spectral diffusion linewidth. As the latter is much smaller than the frequency difference between individual dopants caused by inhomogeneous strain in the crystal, many emitters can be individually addressed and coherently controlled in the same resonator by spectral multiplexing \cite{chen_parallel_2020}.

Among all investigated photon emitters that host long-lived qubits \cite{wolfowicz_quantum_2021, reiserer_cavity-enhanced_2022}, erbium is the only one that exhibits a coherent \cite{bottger_optical_2006} optical transition in the minimal-loss band of fiber-optical telecommunication, around $\SI{1536}{\nano\meter}$. Together with the demonstrated second-long ground-state coherence in high magnetic fields \cite{rancic_coherence_2018}, this is a unique advantage toward the realization of global quantum networks. In recent experiments, single erbium dopants have been resolved in a nanophotonic resonator \cite{dibos_atomic_2018}, and up to four dopants have been controlled simultaneously \cite{chen_parallel_2020}. However, in these experiments the emitters were close to an interface. Thus, fast dephasing has broadened the emission linewidth to $> \SI{10}{\mega\hertz}$, even at a temperature of $\SI{0.5}{\kelvin}$. This has hampered the optical control of a larger number of qubits as well as the optical entanglement of erbium dopants. We demonstrate that this obstacle is overcome in our approach.

\section{Results}

\subsection{Single dopant spectroscopy}

\begin{figure}[tb]
\includegraphics[width=1.0 \columnwidth]{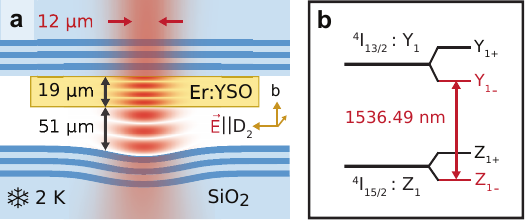}
\caption{\label{fig_Setup}
\textbf{Setup and level scheme}. a) Fabry-Perot cavity (not to scale). A resonator with small mode volume is formed by a flat and a concave glass mirror (blue, bottom and top) with deposited Bragg reflectors (dark blue), confining a stable optical mode (red). Erbium dopants are integrated in a thin, atomically flat crystalline membrane (yellow rectangle). b) Energy level scheme. The studied transition (red) is between the lowest spin and crystal-field levels of the $15/2$ and $13/2$ manifolds of erbium in YSO.
}
\end{figure}

\begin{figure*}[ht]\centering
\includegraphics[width=2.0 \columnwidth]{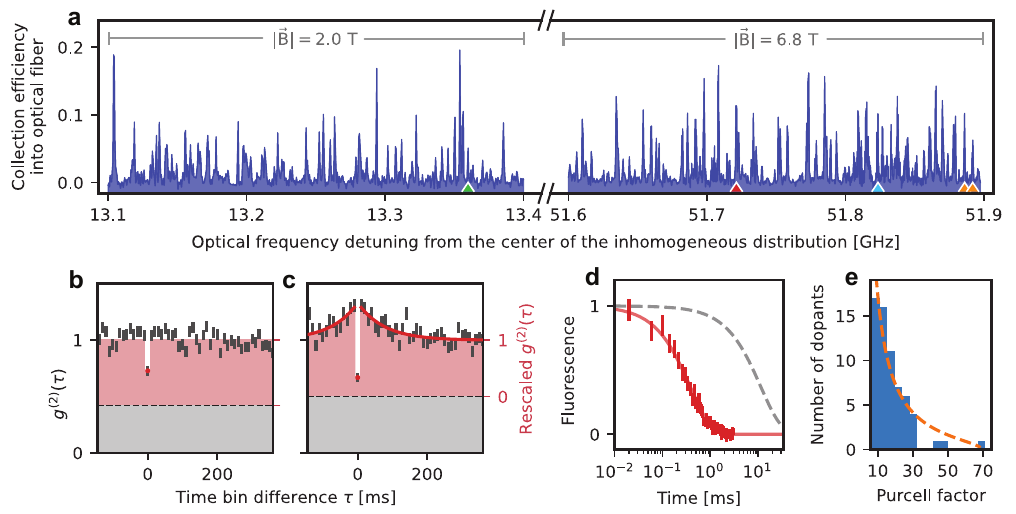}
\caption{\label{fig_broadScan_antibunching}
\textbf{Cavity-enhanced resonance fluorescence spectroscopy}. The error bars denote the $1\sigma$ statistical uncertainty. a) Fluorescence spectrum. The resonator is optically excited with chirped laser pulses of $\SI{0.5}{\mega\hertz}$ bandwidth, generated by an acousto-optical modulator to achieve a large inversion probability. With a repetition rate of $< \SI{2}{\kilo\hertz}$, the fluorescence is measured as a function of the detuning of the pulse center frequency from the center of the inhomogeneous distribution. Over a large frequency range, individual emitters lead to distinct narrow peaks in the spectrum, whose amplitude reflects the strength of the emitter-cavity coupling. b) and c) Photon autocorrelation function $g^{(2)}(\tau)$ measured on a randomly-chosen peak (red triangle in a); raw data (grey) and background-corrected data (red) with subtracted detector dark count contribution (dashed line). b) For fixed-frequency Gaussian excitation pulses with $\SI{0.55}{\mega\hertz}$ (FWHM) bandwidth, a rescaled $g^{(2)}(0)<0.5$ indicates that a peak in the spectrum originates from a single emitter. c) When the excitation bandwidth is reduced to $\SI{0.28}{\mega\hertz}$, the smaller contribution of weakly-coupled dopants gives a reduced coincidence probability at zero delay, approaching the dark count level (dashed). The temporal decay of the bunching observed for short time differences is used to determine the timescale of spectral diffusion, $\SI{80+-20}{\milli\second}$ (red exponential fit curve). d) Purcell enhancement of a randomly-chosen single dopant (red triangle in a). The data and fit (red) indicate a \SI{30.0+-0.8}-fold reduction of the lifetime as compared to dopants in bulk YSO (grey dashed). e) Histogram of the extracted Purcell factors (blue bars) for all prominent peaks between \SI{51.6}{\giga\hertz} and \SI{51.9}{\giga\hertz} in panel a, and theoretical expectation (orange dashed).
}
\end{figure*}

The experimental setup of our Fabry-Perot resonator is shown in Fig. \ref{fig_Setup}a. It is described in detail in \cite{merkel_coherent_2020} and in the Materials and Methods section. The emission of the cavity-coupled erbium dopants is excited by faint laser pulses (that typically contain $\lesssim 10^4$ photons), resonant with the $\SI{1536.49}{\nano\meter}$ transition between the lowest crystal field levels of erbium in YSO, and polarized along the $D_2$ axis, as shown in Fig. \ref{fig_Setup}b. The resonant fluorescence is observed using a superconducting nanowire single-photon detector with a quantum efficiency of $20(2)\,\%$ at a dark count rate of  $\SI{5(2)}{\hertz}$. When scanning the excitation laser frequency and the cavity resonance in parallel, we observe many peaks that originate from single dopants located close to the antinodes of the standing wave cavity field, see Fig. \ref{fig_broadScan_antibunching}a. Their emission is enhanced via the Purcell effect \cite{reiserer_cavity-based_2015, lodahl_interfacing_2015, janitz_cavity_2020} with enhancement factors up to $P=70(12)$, depending on the dopant position in the cavity mode. Such $P$ values, determined from lifetime measurements (panel d and e), are in excellent agreement with the expectation based on the resonator geometry (see Methods). Assuming a random spatial distribution of emitters in the mode, and using the dopant density as the only free parameter, the expected distribution \cite{merkel_coherent_2020} of Purcell factors (orange dashed) shows good agreement with the data (blue bars), except that fewer dopants with the highest Purcell factors are observed. We attribute this to a slight detuning between dopants and cavity in this measurement. Dopants with $P\lesssim 10$ are not included because of the finite signal-to-noise achieved with the used detectors.

For $P\gg 1$, the erbium emission is almost fully channeled into the resonator, enabling high photon generation efficiency. In our setup, the probability that a photon leaves through the coupling mirror is $\SI{34+-3}{\percent}$, and $\SI{63+-9}{\percent}$ of these photons are coupled into a single-mode optical fiber. The overall probability to detect an emitted photon in our setup is further reduced to $\sim2.4\,\%$ by the finite transmission of the fiber-optical setup of $\SI{57+-4}{\percent}$ and the detector efficiency. Each of these values may be further improved in the future.

As a single dopant can only emit one photon at a time, on each peak we observe antibunching when measuring the photon temporal correlation function $g^{(2)}(\tau)$ \cite{reiserer_cavity-based_2015}. The latter is measured using a single device, as the dead time of the detector ($< \SI{0.1}{\micro\second}$) is small compared to the emitter decay ($ > \SI{150}{\micro\second}$). This reduces the dark-count contribution but implies $g^{(2)}(\tau)$=$g^{2}(-\tau)$. In Fig. \ref{fig_broadScan_antibunching}b and c, we exemplarily show the data for a randomly-selected well-coupled dopant with an integration window of $\SI{500}{\micro\second}$, measured at a repetition rate of $\SI{100}{\hertz}$. For a perfect emitter and setup, $g^{(2)}(0)=0$. In the depicted measurement, we observe a finite value of $0.73(4)$, which is in parts explained by dark counts of the detector (dashed line) that could be almost completely eliminated with state-of-the-art superconducting nanowire single-photon detectors that obtain $\SI{0.1}{\hertz}$ dark counts at almost $100\,\%$ quantum efficiency \cite{akhlaghi_waveguide_2015}. In our experiment, we subtract the independently measured dark count contribution and rescale the data \cite{becher_nonclassical_2001}, finding $g^{(2)}(0)=0.53(7)$, limited by background emission from other dopants which are close in emission frequency, but only weakly coupled to the resonator mode because they are located at its side, or at a node of the standing-wave field. Compared to experiments with nanophotonic resonators \cite{chen_parallel_2020}, the increased background originates from the almost thousandfold larger spectral density of emitters in our experiment. It could thus be reduced or avoided by using crystals of lower dopant concentration, potentially even doped only at specific locations. As a straightforward alternative, we reduce the background contribution using excitation pulses of smaller spectral width, e.g. $\SI{0.28}{\mega\hertz}$ in Fig. \ref{fig_broadScan_antibunching}c. Reducing the bandwidth by a factor of two leads to an approximately two-fold reduction of the background, which results in $g^{(2)}(0)=0.34(8)$. This testifies that we indeed observe a single emitter. However, when using narrowband pulses, the excitation probability is diminished if the laser frequency does not precisely match that of the emitter. The latter is not perfectly stable due to the fluctuating magnetic field caused mainly by the nuclear spin bath in YSO. This leads to the observation of bunching when the excitation pulse bandwidth approaches or is reduced below the emitter spectral diffusion linewidth, which will be characterized in more detail below. We observe a characteristic decay constant of $\SI{80+-20}{\milli\second}$ at large magnetic fields, here $\SI{6.8}{\tesla}$.

Thus, the change of the resonance frequency of the dopants happens on a timescale that is slow compared to their lifetime ($\lesssim \SI{0.5}{\milli\second}$). With improved count rates, by using fast resonance frequency measurements and feedback (e.g. via Stark-shifting electrodes \cite{jiri_minar_electric_2009, ruf_quantum_2021, babin_fabrication_2022}), our setup can therefore generate single photons with very narrow linewidth for application in extremely-dense wavelength division multiplexing. The limit of the current measurement is the pulse bandwidth of $\SI{0.28}{\mega\hertz}$, while the ultimate limit is given by the radiative linewidth of $\SI{1.1}{\kilo\hertz}$ FWHM (for the strongest coupled dopants), or by the emitter dephasing. To quantify the latter, we perform coherent spectroscopy.

\subsection{Spectrally multiplexed coherent control of individual dopants}

\begin{figure}[tb]\centering
\includegraphics[width=1.0 \columnwidth]{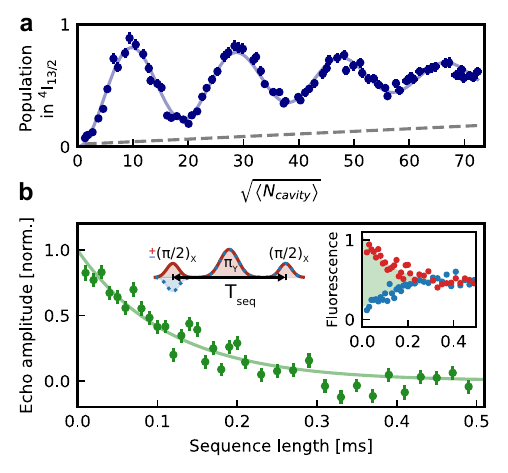}
\caption{\label{fig_RabiOscillation}
\textbf{Optical coherence}. a) An excitation laser pulse of varying intra-cavity mean photon number $\langle N_\text{cavity} \rangle$ is applied to excite the fluorescence of a single dopant. The observation of Rabi oscillations proofs coherent control. The observed decay is well fit by a model (blue line) that includes cavity resonance fluctuations and the excitation of a background of weakly-coupled dopants (grey dashed line). b) Optical spin echo spectroscopy. To measure the optical coherence time, a laser pulse that performs a $\pi/2$ rotation is applied to the dopant. After a delay time of $T_{\mathit{seq}}/2$ (left inset), a $\pi$ pulse cancels the effect of a static detuning between dopant and laser. Finally, another $\pi/2$ pulse is applied. The decay of the difference signal (main panel, green data and exponential fit) between measurements with unchanged (right inset, red) and inverted (blue) phase of the first pulse gives an optical coherence time of $\SI{0.11+-0.01}{\milli\second}$. The error bars denote the $1\sigma$ statistical uncertainty.
}
\end{figure}

To this end, we first establish coherent control over individual dopants by applying laser pulses of constant frequency on resonance with one of the peaks in the spectrum. When the intensity of the Gaussian excitation pulses of $\SI{1}{\micro\second}$ duration (FWHM) is scanned, the fluorescence signal after the pulse shows coherent Rabi oscillations on all investigated dopants, which demonstrates the ability to selectively initialize, control and read the state of individual dopants via frequency multiplexing. In Fig. \ref{fig_RabiOscillation}a, we again show exemplary data of a randomly-chosen emitter (blue triangle in Fig. \ref{fig_broadScan_antibunching}a). We attribute the observed damping to a fluctuation of the pulse intensity which is caused by fluctuations of the cavity resonance frequency between repetitions of the experiment, with a typical oscillation timescale of $\SI{0.1}{\milli\second}$ and an amplitude up to $\SI{6}{\mega\hertz}$ (FWHM) in the used closed-cycle cryocooler. Furthermore, also the background of weakly-coupled dopants contributes to the signal, leading to a linear increase on the observed timescale (grey dashed).

In addition, dephasing may contribute to the damping of Rabi oscillations. To study this, we perform optical spin echo measurements. First, a coherent superposition of the $Z_{1-}$ ground and the $Y_{1-}$ optically excited state is generated by a $\pm\pi/2$-pulse. Then, a $\pi$-pulse cancels the effect of a static detuning between the emitter and the excitation pulse. Finally, another $\pi/2$-pulse transfers the dopants to the ground or optically excited state, depending on the relative phase. Dephasing will reduce the contrast, i.e. the fluorescence signal difference between measurements with orthogonal phase of the two applied $\pi/2$ pulses. In Fig. \ref{fig_RabiOscillation}b, we again show exemplary data from a randomly-chosen dopant (green triangle in Fig. \ref{fig_broadScan_antibunching}a).

The dephasing times of all eight different emitters we measured (four each at $\SI{2}{\tesla}$ and $\SI{6.8}{\tesla}$) are identical within errors, with an average of $T_2=\SI{0.115+-0.007}{\milli\second}$. The corresponding homogeneous linewidth of $\SI{2.3}{\kilo\hertz}$ improves by a factor of five with respect to the narrowest single emitter linewidth reported so far \cite{zhong_optically_2018}. Compared to ensemble measurements in the same setup \cite{merkel_coherent_2020}, the coherence time is slightly reduced.As Er:YSO exhibits a first-order Stark effect, we would not expect that this is an effect of operating at a larger detuning from the center of the inhomogeneous line. Instead, we attribute the broadening to a slightly increased sample temperature caused by the resonator stabilization laser. At lower temperature, lifetime-limited homogeneous linewidth can be achieved at the used resonator parameters \cite{merkel_coherent_2020}.

\subsection{Long-term spectral stability}

\begin{figure}[tb]\centering
\includegraphics[width=1.0 \columnwidth]{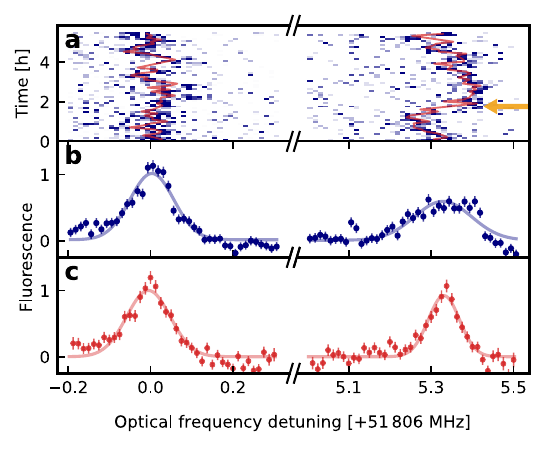}
\caption{\label{fig_TwoIons}
\textbf{Long-term spectral stability}. Two ions, on resonance with the cavity, are optically excited in an alternating sequence with Gaussian pulses of $\SI{0.02}{\mega\hertz}$ FWHM spectral width. The fluorescence is measured as a function of the excitation laser detuning. a) When the results are averaged over $\SI{6}{\minute}$ intervals, the center of the line (red fit results) shows only small fluctuations. Occasional shifts of the center frequency exceeding $\SI{0.1}{\mega\hertz}$ (orange arrow) can be attributed to flips of proximal $^{29}\text{Si}$ nuclear spins. b) When averaging over the full $\SI{6}{\hour}$ measurement record, the Gaussian peaks of the fluorescence exhibit a FWHM of $\SI{0.14+-0.01}{\mega\hertz}$ and $\SI{0.16+-0.02}{\mega\hertz}$. c) When applying feed-forward, i.e. shifting the detuning axis according to the fit center of the last interval to correct for slow frequency shifts, a further linewidth reduction to $\SI{0.10+-0.01}{\mega\hertz}$ (red data and fit) is obtained, depending on the dopant. The error bars denote the $1\sigma$ statistical uncertainty.}
\end{figure}

After characterizing the short-term stability of the emitters, we now turn to long-term observations. None of the measured dopants exhibits any signatures of charge instability or blinking that is commonly observed with many other solid state emitters \cite{ruf_quantum_2021, wolfowicz_quantum_2021}. In Fig. \ref{fig_TwoIons}a, we again show exemplary data of two well-coupled dopants (orange triangles in Fig. \ref{fig_broadScan_antibunching}a) that are detuned by $\SI{5.3}{\mega\hertz}$ and measured in an alternating sequence without cavity length changes. Only small fluctuations of the emitter frequency are observed between 6-minute measurement intervals. The spectral diffusion of both dopants is not correlated, meaning that its source is local to each emitter. We find a standard deviation of the fitted line centers (red) of $\SI{45}{\kilo\hertz}$. Averaging over time, Fig. \ref{fig_TwoIons}b, we obtain Gaussian lines with $\SI{0.14+-0.01}{\mega\hertz}$ and $\SI{0.16+-0.02}{\mega\hertz}$ FWHM (blue).

We performed such long-term measurements on a random sample of six dopants. Over the course of several days, they all give similar temporal behavior to that shown in Fig. \ref{fig_TwoIons}b. Averaging the measurements, we determine a spectral-diffusion linewidth of $\Gamma_{SD}=\SI{0.152+-0.008}{\mega\hertz}$ FWHM. This corresponds to an optical Ramsey decay time of $T_2^* = \frac{1}{\pi \Gamma_{SD}} \simeq \SI{2}{\micro\second}$.

The measured linewidth is much more narrow than that of any previously observed single emitter in any crystalline host \cite{faraon_resonant_2011, lodahl_interfacing_2015, riedel_deterministic_2017, bhaskar_experimental_2020, xia_tunable_2022, wolfowicz_quantum_2021}. Compared to erbium in the same material in the proximity of interfaces \cite{dibos_atomic_2018, chen_parallel_2020}, we find an improvement by two orders of magnitude. Compared to other single rare-earth dopants in tailored hosts with no first-order Stark shift \cite{zhong_optically_2018, kindem_control_2020}, a tenfold narrower line is achieved. Taken together, these results suggest that our approach of avoiding the proximity to interfaces eliminates the main source of spectral instability encountered in nanophotonic devices.

The observed spectral diffusion linewidth is in good agreement with the expectation from the coupling to the fluctuating bath of yttrium nuclear spins within a few nm from each dopant. We calculate this to be $\SI{0.20}{\mega\hertz}$ assuming purely dipolar interactions, which is justified in a large magnetic field, as detailed in \cite{merkel_enhancing_2021}. Thus, a further reduction of the spectral diffusion may be expected with hosts that contain fewer nuclear spins \cite{weiss_erbium_2021, gritsch_narrow_2022, le_dantec_twenty-threemillisecond_2021, stevenson_erbium-implanted_2022}.

In the current samples, the absence of other, dominant sources of frequency instability enables the exploration of the dynamics of the frozen core of nuclear spins surrounding individual emitters \cite{wolfowicz_quantum_2021}. As an example, we observe that some dopants (e.g. the right peak in Fig. \ref{fig_TwoIons}) show occasional fast shifts of the center frequency exceeding $\SI{0.1}{\mega\hertz}$ (orange arrow in Fig. \ref{fig_TwoIons}a). We attribute these shifts to flips of proximal $^{29}\text{Si}$ nuclear spins that have a natural isotopic abundance of $\sim 5\,\%$ and an expected lifetime that is consistent with the observed absence of such shifts over several hours \cite{bottger_optical_2006}. Remarkably, the maximum expected $^{29}\text{Si-Er}$ coupling of $\SI{0.36}{\mega\hertz}$ can exceed the measured spectral-diffusion linewidth. Thus, our system may enable quantum network nodes with optically-interfaced nuclear-spin registers. In turn, undesired nuclear-spin induced shifts may be avoided in isotopically purified samples, or by applying a feed-forward operation. To implement the latter, we adapt the detuning of the excitation laser based on the center-frequency fit in the last measurement interval, achieving a further linewidth reduction below $\SI{0.10}{\mega\hertz}$ (in post-processing) for the dopant shown in Fig. \ref{fig_TwoIons}c. Using this technique to further reduce the linewidth of all dopants would require faster resonance frequency measurements, ideally within the observed spectral diffusion timescale of $\SI{80+-20}{\milli\second}$ (see Fig. \ref{fig_broadScan_antibunching}c). This may become feasible with further improvements of the experimental setup.

\section{Discussion}

In summary, we have shown that the fluorescence of individual emitters in a solid can be enhanced without inducing significant spectral instability by proximal interfaces. The resulting ultra-narrow spectral diffusion linewidths enable cavity-enhanced quantum network nodes at a telecommunication wavelength in which hundreds of emitters can be individually controlled via ultra-dense wavelength-division multiplexing. While our approach is based on erbium dopants in YSO, it can be readily transferred to many other emitters in host crystals that can be polished, etched or grown to form smooth membranes of a few micron thickness; the most prominent ones are diamond \cite{riedel_deterministic_2017, ruf_optically_2019}, silicon, silicon carbide \cite{song_ultrahigh-q_2019, lukin_4h-silicon-carbide--insulator_2019} and a large variety of other semiconductors \cite{rogers_synthesis_2011}, i.e. the majority of single-emitter hosts studied so far \cite{wolfowicz_quantum_2021}.

In our approach, not only the emitter- but also the resonator linewidth is narrow. This will enable single-shot spin-readout without a cycling transition via frequency-selective fluorescence enhancement \cite{reiserer_cavity-based_2015} at low magnetic fields, overcoming the need for emitter-specific, precise field alignment \cite{raha_optical_2020,  chen_parallel_2020}. Still, using the built-in piezo to control the cavity resonance, which has a tuning range of several hundred GHz, one can switch between all optically resolved dopants, and thus increase the number of multiplexed qubits per cavity up to the limit imposed by the inhomogeneous linewidth. Currently, switching is possible on a sub-millisecond timescale \cite{casabone_dynamic_2021}, which may be further reduced by including electro-optic crystalline layers as host crystal \cite{xia_tunable_2022} or additional tuning element. In addition, using techniques of optimal control \cite{werschnik_quantum_2007} may allow for exciting one emitter without exciting others that are close-by in frequency, but differ in coupling strength. This can reduce the two-photon component measured in the correlation function of Fig. \ref{fig_broadScan_antibunching}, and may further increase the multiplexing capability of our setup, as it would alleviate the apparent tradeoff that increasing the number of multiplexed emitters reduces the fidelity of single-emitter control.

The narrow homogeneous linewidth, slow spectral diffusion, and high photon generation efficiency of our setup should enable the entanglement of dopants over tens of kilometers of optical fiber via photon interference \cite{ruf_quantum_2021} or other, cavity-based protocols that can be less sensitive to slowly-fluctuating emitter detuning \cite{reiserer_cavity-enhanced_2022}. The achieved Purcell enhancement reduces the optical lifetime to $\lesssim \SI{0.15}{\milli\second}$. Thus, after $\sim \SI{30}{\kilo\meter}$ the achievable entanglement rate will not be limited by the time it takes to generate a photon, but by the time the photon travels in the fiber. To ensure sufficient ground-state coherence in such setting, one can use hyperfine states of the isotope $^{167}\text{Er}$ \cite{rancic_coherence_2018} or host crystals with a low concentration of nuclear spins and erbium impurities \cite{weiss_erbium_2021, gritsch_narrow_2022, le_dantec_twenty-threemillisecond_2021, stevenson_erbium-implanted_2022} in order to avoid dephasing via spin-spin interactions \cite{merkel_dynamical_2021}. 

Over shorter distances, the rate of optical spin-spin entanglement would however be limited by the emitter lifetime. This may be further reduced using silicon \cite{weiss_erbium_2021, gritsch_narrow_2022} or other \cite{stevenson_erbium-implanted_2022} host crystals, as well as by using Fabry-Perot resonators with higher finesse and smaller mode waist \cite{wachter_silicon_2019, najer_gated_2019}. Combining the two, Fourier-limited spectral diffusion linewidth should be achievable, eliminating the need for fast resonance frequency measurements common in solid-state quantum network nodes \cite{ruf_quantum_2021, reiserer_cavity-enhanced_2022}. Finally, our setup may allow further steps towards distributed quantum information processing with all-to-all connectivity in a rare-earth based quantum computer \cite{kinos_roadmap_2021}.

\section{Materials and Methods}
YSO is a commonly used material for the integration of rare-earth dopants towards quantum applications, as it is commercially available with different dopants in a high crystalline quality and its constituents have small nuclear magnetic moments. The Kramers' dopant erbium \cite{wolfowicz_quantum_2021} substitutes yttrium in two crystallographic sites. Our study is performed using site 1 that has a transition wavelength of $\SI{1536.5}{\nano\meter}$ and a $\SI{414+-7}{\mega\hertz}$ full-width-at-half-maximum (FHWM) Lorentzian inhomogeneous linewidth \cite{merkel_coherent_2020} caused by inhomogeneous strain. We use a nominally undoped YSO membrane of $19(1)\,\si{\micro\metre}$ thickness that contains $<1\,\text{ppm}$ trace impurities of erbium. Even at this low concentration, at the center of the inhomogeneous line the spectral density of erbium is too high to resolve individual dopants. We therefore operate at several $\si{\giga\hertz}$ detuning, i.e. at a frequency where only few erbium dopants are resonant. 

The samples are cooled to $<\SI{2}{\kelvin}$ in a closed-cycle cryo\-cooler. We apply an external magnetic field along the $b$ axis of the crystal, such that the magnetically inequivalent classes are degenerate with a large effective ground state $g$-factor of 9 \cite{bottger_optical_2006}. The field strength is set to $2.0$ or $\SI{6.8}{\tesla}$. As the splitting between the $Z_1$ and $Z_2$ crystal field levels ($>\SI{1}{\tera\hertz}$) and the Zeeman states of the $Z_1$ manifold ($0.3$ or $\SI{0.9}{\tera\hertz}$) is much larger than $k T \simeq \SI{0.04}{\tera\hertz}$, only the lowest Zeeman level of $Z_1$ will be occupied. In addition, most paramagnetic impurities are frozen to the ground state, which reduces magnetic noise from electronic spin flips.

To enhance the emission via the Purcell effect, we use a plano-concave Fabry-Perot resonator whose length can be tuned and stabilized to $\pm \SI{1}{\pico\metre}$ using a piezo tube. To this end, we irradiate a laser at $1593\,\si{\nano\metre}$, which is far detuned from the erbium transition but resonant with another longitudinal cavity mode. Depending on its power, the stabilization laser can also lead to a considerable temperature increase ($\sim \SI{1}{\kelvin}$) of the crystal with respect to the cryostat temperature \cite{merkel_coherent_2020}.

The device fabrication is described in detail in the supporting information of \cite{merkel_coherent_2020}. The radius of curvature of the concave mirror, c.f. Fig. \ref{fig_Setup}a, is $\SI{155+-3}{\micro\metre}$. The mirror transmissions are $22(8)\,\text{ppm}$ (for the flat outcoupling mirror with the bonded crystal) and $20(7)\,\text{ppm}$ (for the concave mirror), which is comparable to the absorption and scattering losses, $27(14)\,\text{ppm}$. This leads to a finesse of $9.0(7)\cdot 10^4$ and a linewidth of $\SI{13+-1}{\mega\hertz}$. From the independently measured mirror parameters we expect a Purcell enhancement of $P_\text{TL}=\SI{362+-26}{}$ for a two-level system at the maximum of the cavity field. For the investigated transition, with the polarization of the excitation laser and the emitted photons parallel to the D2 axis of the YSO crystal, the branching into other crystal field levels reduces this value to $P=\SI{74+-7}{}$, in good agreement with the measurements. This value is slightly larger than that of our earlier measurements on erbium ensembles \cite{merkel_coherent_2020} because of the changed polarization, in combination with an increased mirror separation. This reduces $P_\text{TL}$, but increases the branching ratio and thus $P$ \cite{merkel_enhancing_2021}.

\section{Acknowledgements}
This project received funding from the European Research Council (ERC) under the European Union's Horizon 2020 research and innovation programme (grant agreement No 757772), and from the Deutsche Forschungsgemeinschaft (DFG, German Research Foundation) under Germany's Excellence Strategy - EXC-2111 - 390814868. The authors declare that they have no competing interests.

\bibliographystyle{ScienceAdvances}
\bibliography{bibliography.bib}

\end{document}